\documentclass{elsart}

 \usepackage{graphicx}

\usepackage{amssymb}

\begin{document}

\begin{frontmatter}



\title{Angular Pseudomomentum Theory for the Generalized Nonlinear Schr\"{o}dinger Equation in Discrete Rotational Symmetry Media}


\author[impaUPV]{M.-\'A. Garc\'{\i}a-March},
\author[opticsUV]{A. Ferrando},
\author[impaUPV]{M. Zacar\'es},
\author[teorUV]{J. Vijande},
 and
\author[Smines]{L. D. Carr}
\address[opticsUV]{Interdisciplinary Modeling Group, \textit{InterTech}. Departament d'\'Optica, Universitat de Val\`encia, Spain.}
\address[impaUPV]{Interdisciplinary Modeling Group, \textit{InterTech}. Instituto Universitario de Matem\'atica Pura y Aplicada. Universidad Polit\'ecnica de Valencia, Spain.}
\address[teorUV]{Departament de Fisica Te\'orica. Universitat de Val\`encia, Spain.}
\address[Smines]{Department of Physics, Colorado School of Mines, Golden, CO 80401, USA.}

\begin{abstract}
We develop a complete mathematical theory for the symmetrical solutions of the generalized nonlinear Schr\"odinger equation based on the new concept of angular pseudomomentum.
We consider the symmetric solitons  of a generalized nonlinear Schr\"odinger equation with a nonlinearity depending on the modulus of the field.  We provide a rigorous proof of a set of mathematical results justifying that these solitons can be classified according to the irreducible representations of a discrete group.  Then we extend this theory to non-stationary solutions and study the relationship between angular momentum and pseudomomentum.  We illustrate these theoretical results with numerical examples.  Finally, we explore the possibilities of the generalization of the previous framework to the quantum limit.

\end{abstract}

\end{frontmatter}

\section{Introduction}
\label{Intro}

One-dimensional discrete solitons in periodic media have been  proposed both in the framework of nonlinear optics~\cite{ChristodoulidesOL1988} and  that of Bose-Einstein condensates (BECs)~\cite{TrombettoniPRL2001,AbdullaevPRA2001}.  On the other hand, their two-dimensional counterparts  have attracted intense investigational effort in recent years.  In this context, it has been shown that two-dimensional optical discrete solitons can exist in optically induced lattices~\cite{EfremidisPRE2002,YangOL2003}, in nonlinear photonic crystals~\cite{MingaleevPRL2001,XiePRE2003}, and in nonlinear photonic crystal fibers~\cite{FerrandoOE2003}.  In addition, experimental observations of these solitons in one and two dimensions have been reported  in optically induced lattices~\cite{FleisherPRL2003,FleisherNature2003,NeshevOL2003} and in modulated Bessel optical lattices~\cite{FisherOE2006,WangPRL2006}.  In parallel to these findings, discrete soliton vortices have been proposed both in the context of discrete models~\cite{MalomedPRE2001} and in continuous models of optically induced lattices~\cite{YangOL2003,MusslimaniJOSAB2004},  Bessel lattices~\cite{KartashovPRL2005}, and  photonic crystal fibers~\cite{FerrandoOE2004}.  They have been also experimentally observed in optically induced lattices~\cite{NeshevPRL2004,FleisherPRL2004, FleisherOPN2004}. Other discrete solitons, such as dipole and quadropole solitons, have been shown to exist by means of numerical simulations in optically induced lattices~\cite{MusslimaniJOSAB2004,YangOL2004, YangSAM2004}, in Bessel lattices~\cite{KartashovPRE2004}, and in photonic crystal fibers~\cite{FerrandoOE2005}, and they have been observed in the former system~\cite{YangOL2004}.
Bright discrete  solitons, vortex solitons and dipole solitons have been also  proposed in a self-attractive (or focusing) BEC loaded in an optical lattice~\cite{YangOL2003,MusslimaniJOSAB2004,BaizakovEL2003}. Different results have been obtained in the self-repulsive (or defocusing) case. Extended solitons or BEC arrays have been realized in 2D optical lattices~\cite{GreinerPRL2001,GreinerNATURE2002}. These BEC arrays displayed coherent spatial structures in 2D induced by  modulational instability~\cite{BaizakovJPB2002}.  This was the mechanism that permitted researchers to propose in-phase and $\pi$ out-of-phase discrete bright solitons  in self-defocusing media~\cite{OstrovskayaPRL2003,EfremidisPRL2003}.

All these localized structures have been found as soliton solutions of 2D nonlinear Schr\"odinger Equations (NLSE) with a periodic potential and different expressions for the nonlinear part (Kerr focusing or defocusing nonlinearity, saturable, etc) -- we will refer to all such equations as \emph{generalized nonlinear Schr\"odinger Equations}.  In addition, the amplitude of all these solitons has discrete rotational symmetry.  In Ref.~\cite{FerrandoOE2005} it was proven that a soliton of a generalized NLSE in a discrete rotationally symmetric potential with a nonlinearity depending on the modulus of the field must belong to some representation of the corresponding discrete rotational group, provided that its modulus is also discrete rotational invariant.  As proved in Ref.~\cite{FerrandoPRL2005}, this property implies  that there must exist a ``cut-off'' in the allowed values of the vorticity of a discrete vortex soliton in discrete rotational symmetry media, as long as the discrete vortex presents a single singularity.  It was also demonstrated that identical results can be obtained if the discrete vortex soliton is reinterpreted as an angular Bloch mode characterized by its angular pseudomomentum, which is conserved along the direction of propagation~\cite{FerrandoPRE2005}. Another important result is the existence of a rule that links angular momentum and angular pseudomomentum in an interface between a rotationally invariant and a discrete rotationally invariant medium~\cite{FerrandoPRL2005b}. This rule permits the study of vorticity transformations  in this kind of interfaces~\cite{FerrandoPRL2005b,Perez_GarciaPRA07} and some of these transformations have been experimentally observed~\cite{BezryadinaOE2006}.

In this paper we will rigorously prove that the solitons  of a generalized NLSE with a nonlinearity depending on the modulus of the field can be classified according to the irreducible representations of a discrete group, if that modulus is discrete rotationally invariant. We will show that, consequently,  a different quantity can be assigned to discrete solitons, discrete vortex solitons, and discrete $k$-pole solitons (for example, dipole or quadrupole solitons).  We will name this quantity \emph{angular pseudomomentum} to be consistent with the nomenclature of Ref.~\cite{FerrandoPRE2005}. We will also offer detailed proofs of the conservation of angular pseudomomentum and transformation rules between angular momentum and pseudomomentum.  Finally, we will outline the extension of these results to the quantum limit.

The paper is organized as follows. In the Sec.~\ref{Class}, the classification of discrete solitons according to  angular pseudomomentum is demonstrated.  In Sec.~\ref{Solu} examples of this classification are provided.  Section~\ref{Dynam} is devoted to the conservation of angular pseudomomentum and to the transformation rules between angular momentum and pseudomomentum.  In Sec.~\ref{Theor} the extension of the previous results to the quantum limit is outlined. Finally, in Sec.~\ref{Conc} the conclusions are presented.

\section{Classification of Discrete Rotational Symmetric Solitons}
\label{Class}

Let $L_{0}(\mathbf{x})\equiv\triangle-V(\mathbf{x})$, where $\mathbf{x}\equiv(x,y)\in\mathbb{R}^{2}$,  $\triangle\equiv\frac{\partial^{2}}{\partial x^{2}}+\frac{\partial^{2}}{\partial y^{2}}$
and $V(\mathbf{x})$ is a function invariant under a symmetry group $\mathcal{G}$. Let us assume that
$\mathcal{G}$ is the rotational symmetry group
$\mathcal{O}(2)$ (not to be confused with an order of magnitude symbol), a discrete rotational symmetry group $\mathcal{C}_{n}$ of order $n$,
or a  discrete rotation-reflection symmetry group
$\mathcal{C}_{nv}$.

 Let 
$L_{1}(\vert\psi\vert^{2})\equiv F(\vert\psi\vert^{2})$, where  $\psi=\psi(\mathbf{x})$ is a function such that
$\vert\psi\vert^{2}$ is invariant under a symmetry group $\mathcal{G}'$ and
$F$ is a function invariant under the same group as $\vert\psi\vert^{2}$. We will assume that $\mathcal{G}'$, the symmetry group of  $\vert\psi_{\nu}\vert^{2}$, is a subset or coincides with $\mathcal{G}$, the symmetry group of $V(\mathbf{x})$.
Let $L(\mathbf{x},\vert\psi\vert^{2})\equiv L_{0}+L_{1}$. Under these assumptions, let us consider the eigenvalue equation
\begin{equation}
L(\mathbf{x},\vert\psi\vert^{2})\psi=\mu\psi.\label{eq:ec_estacionarias}\end{equation}
A solution $\psi$ of this equation  gives rise to a stationary solution   $\phi(\mathbf{x},t)=e^{i\mu t}\psi(\mathbf{x})$ of the  generalized NLSE
\begin{equation}
L(\mathbf{x},\vert\phi\vert^{2})\phi=-i\frac{\partial \phi }{\partial t }.\label{eq:ec_evolucion}\end{equation}
Let $\psi_{\nu}(\mathbf{x})$ be a solution of Eq. (\ref{eq:ec_estacionarias}).  We call the \textit{differential operator
generated by $\psi_{\nu}$} the function $L(\mathbf{x},\vert\psi_{\nu}\vert^{2})$.
The eigenfunctions $\psi$  of this operator with eigenvalue $\mu$ satisfy
 $L(\mathbf{x},\vert\psi_{\nu}\vert^{2})\psi=\mu\psi$.
The function  $\psi_{\nu}(\mathbf{x})$ is a \textit{self-consistent solution} since  $L(\mathbf{x},\vert\psi_{\nu}\vert^{2})\psi_{\nu}=\mu\psi_{\nu}$. The function $\psi_{\nu}$ is a \textit{symmetric self-consistent solution} if $\mathcal{G}'\subseteq\mathcal{G}$. A function  $\psi_{\eta}$ is a  \textit{non-self-consistent solution} if $L(\mathbf{x},\vert\psi_{\nu}\vert^{2})\psi_{\eta}=\mu\psi_{\eta}$
but $L(\mathbf{x},\vert\psi_{\eta}\vert^{2})\psi_{\eta}\neq \mu\psi_{\eta}$.

Note that the differential operator $L(\mathbf{x},\vert\psi\vert^{2})$
is Hermitian since $L=L^{\dagger}$,
and therefore all its eigenvalues $\mu$ are real, $\mu\in\mathbb{R}$. Moreover,
$L=L^{*}$ since $V(\mathbf{x})$ and $F(\vert\psi\vert^{2})$ are real functions.

For a self-consistent solution $\psi_{\nu}$, if we conjugate the eigenvalue equation (\ref{eq:ec_estacionarias}) we find that $L_{0}\psi_{\nu}^{*}+L_{1}(\vert\psi_{\nu}^{*}\vert^{2})\psi_{\nu}^{*}=\mu^{*}\psi_{\nu}^{*}$, since $V$ is a real function and  $\vert\psi_{\nu}\vert^{2}=\vert\psi_{\nu}^{*}\vert^{2}$, $F$ is a real function, and  therefore,  $L_{1}(\vert\psi_{\nu}\vert^{2})=L_{1}(\vert\psi_{\nu}^{*}\vert^{2})$. Hence, $\psi_{\nu}^{*}$ is also a self-consistent solution with eigenvalue  $\mu^{*}$ of the same operator. Finally, $\mu=\mu^{*}$ due to the hermiticity of $L$. This proves the following Lemma:

\begin{lem}
\label{pro:complejo_conjugado_tb_sol} If $\psi_{\nu}$ is a self-consistent solution with eigenvalue $\mu$, $\psi_{\nu}^{*}$ is also a self-consistent solution with the same eigenvalue.
\end{lem}

Let us prove now the following important Lemma:

\begin{lem}
\label{pro:psi_sol_autocons_y_tb_de_otro_operador_} If $\psi_{\nu}$
is a self-consistent solution with eigenvalue
$\mu$, and $\psi_{\sigma }$ is a function such that $L(\mathbf{x},\vert\psi_{\nu}\vert^{2})\psi_{\sigma}=\mu_{\sigma}\psi_{\sigma}$,
 then $\psi_{\sigma }$  is a self-consistent solution if and only if (i) 
 $\psi_{\nu}=k\psi_{\sigma}$ or, (ii) $\psi_{\nu}=k\psi_{\sigma}^{*}$,
with $k\in\mathbb{C}$ and $\vert k\vert^{2}=1$.
\end{lem}

If $\psi_{\sigma}$ is a self-consistent solution then
$$L(\mathbf{x},\vert\psi_{\sigma}\vert^{2})\psi_{\sigma}=\mu\psi_{\sigma}.$$
By hypothesis
$$L(\mathbf{x},\vert\psi_{\nu}\vert^{2})\psi_{\sigma}=\mu\psi_{\sigma}.$$
Now we substract the  equations above to obtain

\[
F(\vert\psi_{\nu}\vert^{2})\psi_{\sigma}-F(\vert\psi_{\sigma}\vert^{2})\psi_{\sigma}=0,\,\,\,\forall\mathbf{x}.\]
For most  physical systems $F$ is an injective function (i.e., one to one) and therefore $\vert\psi_{\nu}\vert^{2}=\vert\psi_{\sigma}\vert^{2}$, $\forall \mathbf{x}$. This fact implies (i) $\psi_{\nu}=k\psi_{\sigma}$ or, (ii) $\psi_{\nu}=k\psi_{\sigma}^{*}$ with $k\in\mathbb{C}$ and $\vert k\vert=1$.

On the other hand, if (i) $\psi_{\nu}=k\psi_{\sigma}$ or, (ii) $\psi_{\nu}=k\psi_{\sigma}^{*}$,  obviously $\psi_{\sigma}$ is a self-consistent solution since, in both cases, $L(\mathbf{x},\vert\psi_{\sigma}\vert^{2})\psi_{\sigma}=\mu\psi_{\sigma}$. This proves the previous Lemma.

Now we proceed to apply this Lemma to  the transformed functions $\mathcal{T}_{i}'\psi_{\nu}$, where $\mathcal{T}_{i}'$ is  an element of the group $\mathcal{G}'$. Before doing this we observe that

\[
\mathcal{T}_{i}'L(\mathbf{x},\vert\psi_{\nu}\vert^{2})\psi_{\nu}(\mathbf{x})=\mathcal{T}_{i}'\mu_{\nu}\psi_{\nu}(\mathbf{x}),\]
or \[
\mathcal{T}_{i}'L(\mathbf{x},\vert\psi_{\nu}\vert^{2})\mathcal{T}_{i}'^{-1}\mathcal{T}_{i}'\psi_{\nu}(\mathbf{x})=\mu_{\nu}\mathcal{T}_{i}'\psi_{\nu}(\mathbf{x}).\]

It is clear that $L(\mathbf{x},\vert\psi_{\nu}\vert^{2})$ is invariant under the symmetry group  $\mathcal{G}'$ of $\vert\psi_{\nu}\vert^{2}$.  Firstly, $\triangle$ is invariant under the symmetry group $\mathcal{O}(2)$, and consequently it is invariant under any discrete rotational symmetry group. Secondly, the function $V(\mathbf{x})$ is invariant under the symmetry  group  $\mathcal{G}$. Since we have assumed that $\mathcal{G}'\subseteq\mathcal{G}$ it must be also invariant under the symmetry group $\mathcal{G}'$. And finally, $F(\vert\psi_{\nu}\vert^{2})$ is invariant under the symmetry group $\mathcal{G}'$ since $\vert\psi_{\nu}\vert^{2}$ is also invariant under this group. So $\mathcal{T}_{i}'L(\mathbf{x},\vert\psi_{\nu}\vert^{2})\mathcal{T}_{i}'^{-1}=L(\mathbf{x},\vert\psi_{\nu}\vert^{2})$ and therefore 

\[
L(\mathbf{x},\vert\psi_{\nu}\vert^{2})\mathcal{T}_{i}'\psi_{\nu}(\mathbf{x})=\mu\mathcal{T}_{i}'\psi_{\nu}(\mathbf{x}).\]

Hence $\mathcal{T}_{i}'\psi_{\nu}(\mathbf{x})$ is also an eigenfunction
with the same eigenvalue. Then, the conditions of Lemma \ref{pro:psi_sol_autocons_y_tb_de_otro_operador_} are satisfied for $ \psi_{\sigma} =\mathcal{T}_{i}'\psi_{\nu}(\mathbf{x})$.

We conclude that  $\psi_{\sigma}$ is a self-consistent solution verifying
\[
L(\mathbf{x},\vert\psi_{\sigma}\vert^{2})\psi_{\sigma}=\mu\psi_{\sigma}.\]

if and only if $ \psi_{\sigma} =\mathcal{T}_{i}'\psi_{\nu}=k\psi_{\nu}$ or $ \psi_{\sigma} =\mathcal{T}_{i}'\psi_{\nu}=k\psi_{\nu}^*$  with $\vert k\vert=1$.

\begin{prop}
\label{thm:Tipsi_prop_psi_o_psi_ast} Let $\psi_{\nu}(\mathbf{x})$ be
a symmetric self-consistent solution of Eq. (\ref{eq:ec_estacionarias}). Then, for any element $\mathcal{T}_{i}'$ of the group  $\mathcal{G}'$,
 the function $\mathcal{T}_{i}'\psi_{\nu}$
can be only a self-consistent solution of Eq. (\ref{eq:ec_estacionarias}) if and only if it is proportional to $\psi_{\nu}$ or to $\psi_{\nu}^{*}$, up to a complex number of modulus unity.
\end{prop}

We have used that $\mathcal{T}_{i}'L(\mathbf{x},\vert\psi_{\nu}\vert^{2})\mathcal{T}_{i}'^{-1}=L(\mathbf{x},\vert\psi_{\nu}\vert^{2})$ in the demonstration of the previous proposition. This is equivalent to the condition
$$[L,\mathcal{T}_{i}']=0.$$
It is well known in quantum mechanics that if $\mathcal{T}_{i}'$ and $L$ commute, we can have simultaneous eigenfunctions of both operators. This is equivalent to the statement that the eigenfunctions of $L$ characterized by the eigenvalue $\mu$ belong to the irreducible representations of the group $\mathcal{G}'$\cite{Hamermesh}. The representations of $\mathcal{C}_n$ or  $\mathcal{C}_{nv}$ are one or two-dimensional\footnote{All the irreducible representations of a discrete rotational group  $\mathcal{C}_{n}$  are one-dimensional. Since $L=L^*$,  there are pairs of irreducible representations whose characters are mutually conjugate and that share the same eigenvalue~\cite{Hamermesh}.
 In quantum mechanics this is due to the time reversal invariance of the Hamiltonian~\cite{Hamermesh}. We will consider these pairs of irreducible representations as a  doubly-degenerate representation.}. In the case of two-dimensional representations, the rotational elements of these groups are given by two-dimensional unitary matrices. In the diagonal basis this matrices are of the form
\begin{eqnarray}
\left(\begin{array}{cc}
\epsilon &0\\
0&\epsilon^*\\
\end{array}
\right)\,,\nonumber
\end{eqnarray}
where $\epsilon=e^{i\frac{2\pi}{n}}$. On the other hand, we have shown that the self-consistent solution $\psi_{\nu}$
verifies that $\mathcal{T}_i'\psi_{\nu}=k\psi_{\nu}$ where $|k|=1$. Therefore, $\psi_{\nu}$ is an eigenfunction of $\mathcal {T}_i'$ with eigenvalue $k$. Comparison with the previous expression implies that $k=\epsilon$. Moreover, group theory insures that the other eigenfunction of the  $\mathcal {T}_i'$ operator has to be $\psi_{\nu}^*$~\cite{Hamermesh}.
We can summarize these arguments in the following Theorem:
\begin{thm}
\label{thm:clasificacion_soluciones_simetricas} Let $\psi_{\nu}=\psi_{\nu}(\mathbf{x})$
be a symmetric self-consistent solution
and let us assume that $\mathcal{G}'\subseteq\mathcal{G}$. Then $\psi_{\nu}$
belongs to some of the irreducible representations of $\mathcal{G}'$.
\end{thm}

But, moreover, we have proved also the following Theorem:

\begin{thm}
\label{thm:clasificacion_soluciones_simetricas_fuerte}
Under the conditions of the previous Theorem and
for  two-dimen\-sio\-nal representations, the only self-consistent solutions are those that form the diagonal basis, i.e., the pair  $\psi_{\nu}$ and $\psi_{\nu}^*$ that fulfills, on the one hand,
$L(\mathbf{x},\vert\psi_{\nu}\vert^{2})\psi_{\nu}=\mu\psi_{\nu}   $ and $L(\mathbf{x},\vert\psi_{\nu}^*\vert^{2})\psi_{\nu}^*=\mu\psi_{\nu}^*    $, and, on the other hand,  $\mathcal {T}_i'\psi_{\nu}=\epsilon\psi_{\nu}$ and $\mathcal {T}_i'\psi_{\nu}^*=\epsilon^*\psi_{\nu}^*$.
\end{thm}

\begin{rem}
\label{not:soluciones_de_subgrupos_de_V} Note that if $\mathcal{G}'\subset\mathcal{G}$ the symmetric self-consistent solutions belong to irreducible representations of the subgroup  $\mathcal{G}'$. This is the case of the nodal solutions introduced in Ref.~\cite{FerrandoOE2005} or the dipole solutions introduced in \\ Refs.~\cite{MusslimaniJOSAB2004,YangOL2004,YangSAM2004}.
\end{rem}

\begin{rem}
Discrete rotation-reflection groups $\mathcal{C}_{nv}$ present one- and two-dimen\-sio\-nal representations. Under the $p^{th}$ rotational element of the group $\mathcal{C}_{n}^{p}$ the function transforms as  $\mathcal{C}_{n}^{p}\psi=k_{p}\psi$ where $k_{p}\in\mathbb{C}$
is a unitary constant. But under the reflection elements of the group $\Pi_{\nu}$, the function is transformed as
$\Pi_{\nu}\psi=k_{\nu}\psi^{*}$
or as  $\Pi_{\nu}\psi=k_{\nu}\psi$, where $k_{\nu}\in\mathbb{C}$ is a unitary constant. So these transformations could involve the function and its complex conjugate.
\end{rem}

Discrete rotational groups, $\mathcal{C}_{n}$, present  $\frac{n}{2}+1$  irreducible representations for even $n$
and $\frac{n-1}{2}+1$ for odd $n$, if the mutually conjugated representations can be considered as a single two-dimensional representation. Two of these representations are one-dimensional, while the rest are two-dimensional. Adding reflections to $\mathcal{C}_{n}$ to form a $\mathcal{C}_{nv}$ group means, on the one hand, that these mutually conjugated representations  form a single two-dimensional representation. On the other hand, the two one-dimensional representations split due to different behaviours of the functions with respect to the reflection axes of the group~\cite{Hamermesh}. In both kinds of groups, all the transformation rules can be satisfied by functions of the form
 $\psi(r,\theta)=e^{i\theta m}u(r,\theta)$,
where $u(r,\theta)=u(r,\theta+\frac{2\pi}{n})$ and $(r,\theta)$ are polar coordinates. Therefore, $u(r,\theta)$  is some periodic function in $\theta$ with period  $\frac{2\pi}{n}$. Moreover
 $m\in\{0,\pm1,\pm2,\ldots,\frac{n}{2}\}$ for even  $n$ or $m\in\{0,\pm1,\pm2,\ldots,\pm\frac{n-1}{2}\}$
for odd  $n$ according to Th. \ref{thm:cut_off_en_m}. Functions with $m=0$ belong to the first one-dimensional representation of  $\mathcal{C}_{n}$ or to some of the  one-dimensional representations of $\mathcal{C}_{nv}$.  Each of these one-dimensional representations have a different behaviour under reflections.  Functions with $m=\frac{n}{2}$, $n$ even,  belong to the other one-dimensional representation of  $\mathcal{C}_{n}$ or to some of the  one-dimensional representations  of $\mathcal{C}_{nv}$. The rest of the functions with $m\in\{\pm1,\pm2,\ldots,\pm\frac{n}{2}\mp1\}$ for even $n$ or $\{\pm1,\pm2,\ldots,\pm\frac{n-1}{2}\}$ for odd $n$ belong to two-dimensional representations. This can be summarized in the following Theorem:

\begin{thm}
(Cut-off for $m$).\label{thm:cut_off_en_m}
The modulus of $m$
presents a maximum value, i.e.,  $\vert m \vert \leq\frac{n}{2}$ for even $n$ and $\vert m \vert\le\frac{n-1}{2}$
for odd  $n$.
\end{thm}

We will name the variable
$m$ \textit{angular pseudo-momentum}, in accordance with~\cite{FerrandoPRE2005}.
We will call solutions with $m=0$, \emph{fundamental} solitons. We will call solutions with $m=\frac{n}{2}$, for even  $n$,  \emph{nodal} or $k$-pole solitons. Finally, we will call the solutions with  $m$ equal to one of the other allowed values \emph{vortex} solitons.

\section{Classification of Solutions According to their Angular Pseudomomentum in $\mathcal{C}_{4v}$ and $\mathcal{C}_{6v}$}
\label{Solu}

Let us numerically calculate some solutions in $\mathcal{C}_{4v}$ and $\mathcal{C}_{6v}$ with all possible angular pseudomomenta  using a relaxation numerical method. We will calculate only solutions with no more than one singularity and with trivial behaviour under reflections. We leave the study of solutions with more singularities and different behaviour under reflections to future work. We will solve Eq. (\ref{eq:ec_estacionarias}) with $F(\vert\psi\vert^{2})=g(\mathbf{x})\vert\psi\vert^{2}$, where $g(\mathbf{x})=V(\mathbf{x})$, i.e., in the self-focusing case.

\begin{figure}[h]
\begin{tabular}{cc}
 (a)&
\hspace{1.6cm}(b)\tabularnewline
\includegraphics[%
  scale=0.4]{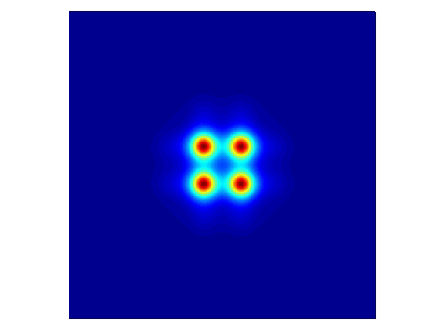}&
\includegraphics[%
  scale=0.4]{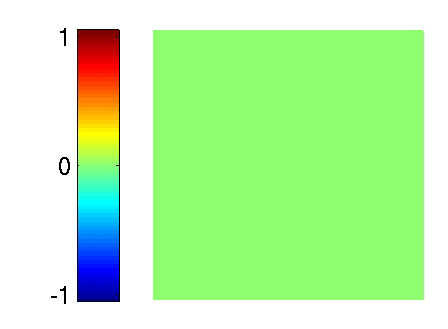} \tabularnewline
(c)&
\hspace{1.6cm}(d)\tabularnewline
\includegraphics[%
  scale=0.4]{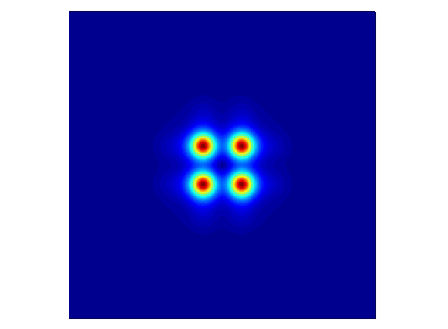}&
\includegraphics[%
  scale=0.4]{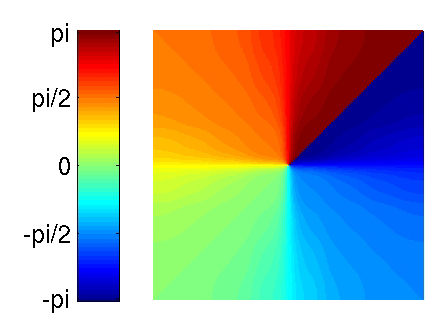} \tabularnewline
(e)&
\hspace{1.6cm}(f)\tabularnewline
\includegraphics[%
  scale=0.4]{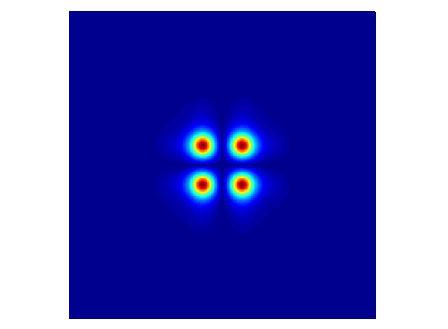}&
\includegraphics[%
  scale=0.4]{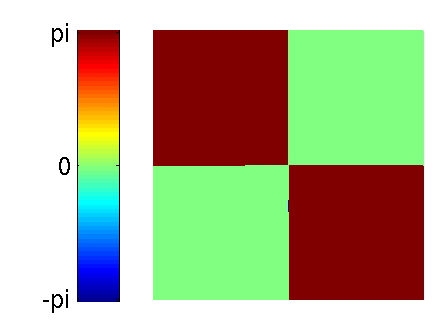} \tabularnewline
\end{tabular}
\caption{\label{cap:Soluciones-simples-C4} Symmetric self-consistent solutions when
$\mathcal{G}$ and $\mathcal{G}'$ are $\mathcal{C}_{4v}$, and  $\mu=1$. The amplitude of the
solution with angular pseudomomentum $m\in\{0,-1,2\}$ is represented in (a), (c), and (e), respectively. In (b), (d), and  (f)
the corresponding phases  of these solutions are represented.}
\end{figure}

\subsection{Solutions in $\mathcal{C}_{4v}$}
\label{c4v}

In this case $V(\mathbf{x})=V_0\left[ \cos^2 (x)+\cos^2 (y) \right]$, with $V_0=2$. This potential can model the propagation of light in an optically induced lattice over a photorefractive crystal or a BEC loaded in an optical lattice in the attractive case. In both cases  appropriate transformations in the physical variables must be done to obtain  Eq. (\ref{eq:ec_evolucion}). We will consider solutions with $\left|\psi\right|^{2}$   invariant under the same group as $V(\mathbf{x})$, i.e., $\mathcal{C}_{4v}$. Therefore, $n=4$  and the allowed values of $m$ according to Th. \ref{thm:cut_off_en_m} are $m\in\{0,\pm1,2\}$.
In Fig. \ref{cap:Soluciones-simples-C4} the amplitude and phase of these solutions are presented.
They are described by the character table of the corresponding irreducible representation since they are correctly transformed, i.e., as $e^{im\varphi}$, under the action of the rotational elements  of the group, i.e., rotations of $\varphi=p\frac{2\pi}{4}$ radians  with $p\in\mathbb{Z}$.
The fundamental soliton ($m=0$) and nodal soliton ($m=2$) solutions belong to one-dimensional representations.  The vortex soliton solution ($m=-1$)  belongs to a two-dimensional irreducible representation; its complex conjugate is also a solution, but with $m=1$.

\begin{figure}

\begin{tabular}{cc}
 (a)&
\hspace{1.6cm}(b)\tabularnewline
\hspace{0.4cm}\includegraphics[%
  scale=0.35]{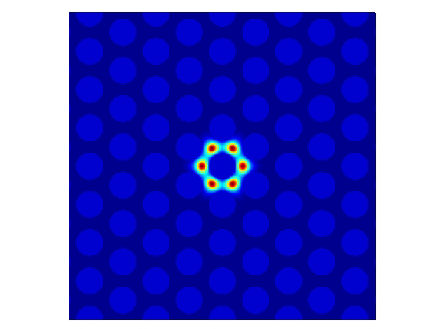}&
\includegraphics[%
  scale=0.35]{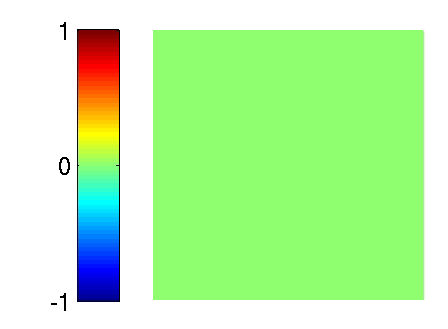} \tabularnewline
(c)&
\hspace{1.6cm}(d)\tabularnewline
\hspace{0.4cm}\includegraphics[%
  scale=0.35]{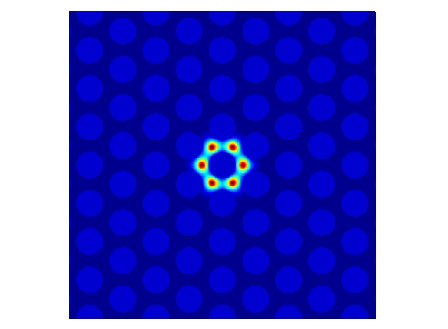}&
\includegraphics[%
  scale=0.35]{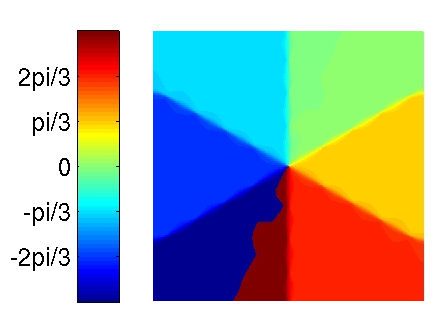} \tabularnewline
(e)&
\hspace{1.6cm}(f)\tabularnewline
\hspace{0.4cm}\includegraphics[%
  scale=0.35]{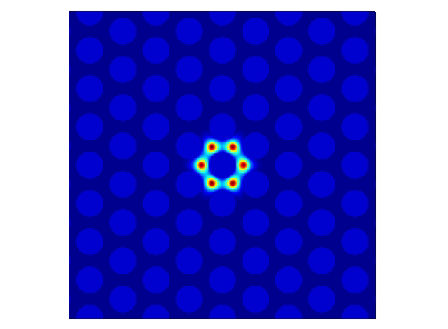}&
\includegraphics[%
  scale=0.35]{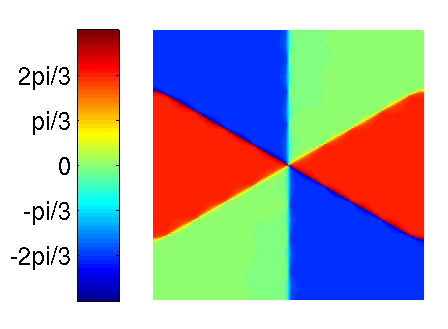} \tabularnewline
(g)&
\hspace{1.6cm}(h)\tabularnewline
\hspace{0.4cm}\includegraphics[%
  scale=0.35]{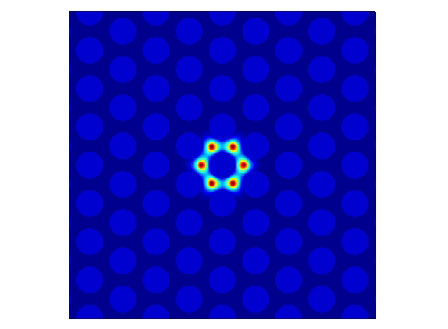}&
\includegraphics[%
  scale=0.35]{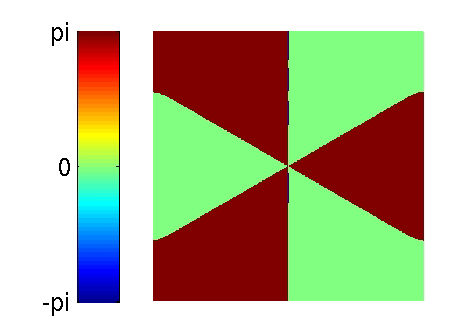} \tabularnewline

\end{tabular}

\caption{\label{cap:Soluciones-simples-C6} Symmetric self-consistent solutions when
$\mathcal{G}$ and $\mathcal{G}'$ are $\mathcal{C}_{6v}$, and  $\mu=-0.06$. The amplitudes of the
solutions with angular pseudomomentum $m\in\{0,-1,-2,3\}$ are represented in (a), (c), (e), and
(g), respectively. The circles in which $V(\mathbf{x})$ is equal to one are represented in light blue. In (b), (d), (f), and (h)
the corresponding phases  of these solutions are represented.}
\end{figure}

\subsection{Solutions in $\mathcal{C}_{6v}$}
\label{c6v}

In this case $V(\mathbf{x})$ is equal to one in circles of radius
 $R=\frac{\bar{R}}{\lambda}$, where $\bar{R}=8000$ and $\lambda=1064$,
and zero outside of these circles. The centers of the circles are located in an
hexagonal grid, i.e., in positions $\mathbf{r}=a[k_{x}+\cos(\alpha)]\mathbf{i}+ak_{y}\sin(\alpha)\mathbf{j}$,
where $\mathbf{i},\mathbf{j}$ define a cartesian coordinate system,  $k_{x},k_{y}\in\mathbb{Z}$, $\alpha=\frac{\pi}{3}$, and $a=\frac{\Lambda}{\lambda}$,
with $\Lambda=23000$. This $V(\mathbf{x})$ is used to model the propagation of light beams in photonic crystal fibers as in Refs.~\cite{FerrandoOE2003,FerrandoOE2004,FerrandoOE2005,FerrandoPRL2005}.

We will consider solutions with $\left|\psi\right|^{2}$   invariant under the same group than $V(\mathbf{x})$, i.e., $\mathcal{C}_{6v}$.
For $n=6$ the allowed values of $m$ acording to Th. \ref{thm:cut_off_en_m} are $m\in\{0,\pm1,\pm2,3\}$.
Fig. \ref{cap:Soluciones-simples-C6} shows the amplitude
and phase of self-consistent solutions with  $m\in\{0,-1,-2,3\}$. It is easy to see that these solutions
fulfill the character table of the corresponding irreducible representation using any rotation of   $p\frac{2\pi}{6}$ rad with $p\in\mathbb{Z}$. For $m=0$ or $m=3$ the solutions are fundamental and nodal solitons, respectively. That is,  they belong to one-dimensional representations of the $\mathcal{C}_{6v}$ group.  For
  $m\in\{-1,-2\}$ the vortex soliton solutions belong to two-dimensional irreducible representations. Their complex conjugates are also solutions but with $m\in\{1,2\}$, respectively.

\section{Dynamics of Rotationally Symmetric Solutions}
\label{Dynam}
\subsection{Conservation of angular pseudomomentum}
\label{Conse}

Let $\phi(\mathbf{x},t_0)$ be a stationary or non-stationary solution of Eq. (\ref{eq:ec_evolucion}) in certain time $t_0$.
Let  its modulus $\left|\phi(\mathbf{x},t_0)\right|^{2}$  be invariant under $\mathcal{C}_{n}$ or $\mathcal{C}_{nv}$, and let us call  its symmetry group $\mathcal{G}'$. As usual, we assume that this group coincides with or is a subgroup of the symmetry group $\mathcal{G}$ of  $V(\mathbf{x})$, where $\mathcal{G}$ can be $\mathcal{O}(2)$, $\mathcal{C}_n$, or  $\mathcal{C}_{nv}$. Using similar arguments as the ones presented in the previous section it can be shown that such a solution must have angular pseudomomentum $m$ and that the value of $m$ must obey Th. \ref{thm:cut_off_en_m}.
Therefore, these solutions are of the form $\phi(\mathbf{x},t_0)=e^{im\theta}u(r,\theta,t_0)$,
with $u(r,\theta+\epsilon,t)=u(r,\theta,t)$ and $\epsilon\equiv\frac{2\pi}{n}$.

If $\phi(\mathbf{x},t_0)$ is a stationary solution, angular pseudomomentum is trivially conserved for $t>t_0$. Let us prove  that if it is not a stationary solution the angular pseudomomentum is also conserved. We will prove this statement by induction. Let us denote
$\phi_0$ by $\phi(\mathbf{x},t_0)$.  It can be shown, using the same arguments as in Sec.~\ref{Class}, that the operator $L(\mathbf{x},\left|\phi_{0}\right|^{2})=L_{0}+L_{1}$
is invariant under the rotational elements  $\mathcal{C}_{\vartheta}$  of the group $\mathcal{G}'$, which must be
$\mathcal{C}_{n}$ or $\mathcal{C}_{nv}$. Therefore, $\mathcal{C}_{\vartheta}L\mathcal{C}_{\vartheta}^{-1}=L$,
where $\vartheta=p\frac{2\pi}{n}$ radians, $p\in\mathbb{Z}$.

Let us discretize the evolution from  $t_{0}=0$ to $t$ using
arbitrarily small steps of size
$h$ with $h\rightarrow0$. Let $t_{j}\equiv t_0+j\,h$, $j\in\{0,1,\ldots\}$, and  $\phi_{j}\equiv\phi(\mathbf{x},t_j)$.
In a first order equation in the evolution variable, the solution $\phi_{j+1}$ can be  obtained in terms of $\phi_{j}$ as $\phi_{j+1}=e^{i\bar{L}(\mathbf{x},\left|\phi_{j}\right|^{2})h}\phi_{j}$.

On the one hand, $\mathcal{C}_{\vartheta}L(\mathbf{x},\left|\phi_{0}\right|^{2})\mathcal{C}_{\vartheta}^{-1}=L(\mathbf{x},\left|\phi_{0}\right|^{2})$ and therefore
$\mathcal{C}_{\vartheta}e^{iL(\mathbf{x},\left|\phi_{0}\right|^{2})h}\mathcal{C}_{\vartheta}^{-1}=e^{iL(\mathbf{x},\left|\phi_{0}\right|^{2})h}$,
since for $h\rightarrow0$ and using the series expansion for the exponential
$\mathcal{C}_{\vartheta}\left(1+iL(\mathbf{x},\left|\phi_{0}\right|^{2})h\right)\mathcal{C}_{\vartheta}^{-1}=1+iL(\mathbf{x},\left|\phi_{0}\right|^{2})h$.  Consequently,  $\mathcal{C}_{\vartheta}\phi_{1}=\mathcal{C}_{\vartheta}e^{iL(\mathbf{x},\left|\phi_{0}\right|^{2})h}\phi_{0}=e^{iL(\mathbf{x},\left|\phi_{0}\right|^{2})h}\mathcal{C}_{\vartheta}\phi_{0}$.
 Taking into account that $\mathcal{C}_{\vartheta}\phi_{0}=e^{im\vartheta}\phi_{0}$
it follows that $\mathcal{C}_{\vartheta}\phi_{1}=e^{iL(\mathbf{x},\left|\phi_{0}\right|^{2})h}e^{im\vartheta}\phi_{0}$.
Consequently,  $\mathcal{C}_{\vartheta}\phi_{1}=e^{im\vartheta}\phi_{1}$. Then $\phi_{1}$ is  a function with the same angular pseudomomentum as $\phi_{0}$.

On the other hand, let us suppose that $\phi_{j}$ is a function with well defined angular pseudomomentum $m$. Then
$\mathcal{C}_{\vartheta}\phi_{j}=e^{im\vartheta}\phi_{j}$ and  $\mathcal{C}_{\vartheta}L(\mathbf{x},\left|\phi_{j}\right|^{2})\mathcal{C}_{\vartheta}^{-1}=L(\mathbf{x},\left|\phi_{j}\right|^{2})$.
Using the discrete evolution equation and these  two equations in the same way as in the previous paragraph, one finds that $\mathcal{C}_{\vartheta}\phi_{j+1}=e^{im\vartheta}\phi_{j+1}$
and therefore $\phi_{j+1}$ and $\phi_{j}$ share the same angular pseudomomentum. In the continous limit $h\rightarrow0$
both results are  true.  This proves the following Theorem:

\begin{thm}
\label{thm:cons_m} Let $\phi(\mathbf{x},t)$ be a solution of Eq. (\ref{eq:ec_evolucion}) and let us assume that $\mathcal{G}$ is $\mathcal{O}(2)$,
$\mathcal{C}_{n}$ or $\mathcal{C}_{nv}$. Let us assume as well that the symmetry group $\mathcal{G}'$ of the solution at certain $t_0$,  $\phi(\mathbf{x},t_{0})$, is $\mathcal{C}_{n}$
or $\mathcal{C}_{nv}$ and $\mathcal{G}'\subseteq\mathcal{G}$. Then the solution presents angular pseudomomentum $m$ and  is conserved for $t>t_0$.
\end{thm}

A solution with well defined  angular momentum must be an eigenfunction
of $\mathcal{R}\phi=e^{il\theta}\phi$,
where $l$ is the angular momentum  and $\mathcal{R}=e^{-i\theta(\frac{\partial}{\partial\theta})}$. If the evolution operator $L$
is invariant under $\mathcal{O}(2)$, $l$ is conserved and  coincides
with the $z$ component of the angular momentum:
\begin{equation}
j_{z}=-i\int_{\mathbb{R}^{2}}dxdy\phi^{*}(r\times\nabla)\phi\,,\label{eq:ang_momentum}
\end{equation}
where  $\phi$ is a normalized eigenfunction of $\mathcal{R}$. In a system described by a nonlinear evolution operator $L$ as defined before,  this is no longer true if the function $\phi(\mathbf{x},t_{0})$ is such that $|\phi(\mathbf{x},t_{0})|^{2}$
is invariant under $\mathcal{C}_{n}$ or $\mathcal{C}_{nv}$,  even in the case where
$\mathcal{G}$ is $\mathcal{O}(2)$. Of course, it is not true if $\mathcal{G}$ is $\mathcal{C}_{n}$ or $\mathcal{C}_{nv}$.

\subsection{Transformation rule}
\label{Trans}

Let us study the solutions of Eq. (\ref{eq:ec_evolucion}) when the solution has a well defined angular momentum for $t  < t_1$  and the symmetry of $V(\mathbf{x},t)$ changes  into a discrete group $\mathcal{C}_{n}$ or $\mathcal{C}_{nv}$ at $t_1$, i.e.:

\begin{equation}
V(r,\theta,t)=\left\{ \begin{array}{ll}
V(r), & \mbox{if $t<t_{1}$}\\
V(r,\theta),\,\,\,|\,\,\, V(r,\theta)=V(r,\theta+\frac{2\pi}{n})\,\, & \mbox{if $t\ge t_{1}.$}\end{array}\right.\label{eq:cambio_V}\end{equation}
The function $\phi(\mathbf{x},t<t_{1})$ has well defined angular momentum and $\phi(\mathbf{x},t=t_{1})$ also has well defined angular pseudomomentum $m$, since any discrete rotational group is a subgroup of  $\mathcal{O}(2)$. Therefore, $\mathcal{C}_{n}^{p}\phi(\mathbf{x},z=z_{1})=e^{im\vartheta}\phi(\mathbf{x},t=t_{1})$
with $\vartheta=p\frac{2\pi}{n}$ radians, where $p\in\mathbb{Z}$ and $\mathcal{C}_{n}^{p}$
is a rotation element of $\mathcal{C}_{n}$
or $\mathcal{C}_{nv}$. This $m$ is conserved for $t>t_1$ due to Th. \ref{thm:cons_m}. Let us see that the final value of $m$ depends on the initial value of $l$ and on the order of symmetry $n$.
Let $\phi_{l}(\mathbf{x},t\le t_{1})$ be the function with angular momentum $l$ and $\phi_{m}(\mathbf{x},t\ge t_{1})$ the function with angular pseudomomentum $m$. The projection coefficient of  $\phi_{l}$ on
$\phi_{m}$ is
$c_{ml}=\int_{\mathbb{R}^{2}}\phi_{m}^{*}(\mathbf{x},t_{1})\phi_{l}(\mathbf{x},t_{1})d^{2}\mathbf{x}$. Also, $\phi_{m}(r,\theta+\frac{2\pi}{n},t_{1})=e^{im\frac{2\pi}{n}}\phi_{m}(r,\theta,t_{1})$
and $\phi_{l}(r,\theta+\frac{2\pi}{n},t_{1})=e^{il\frac{2\pi}{n}}\phi_{l}(r,\theta,t_{1})$.
So if we take the transformation $\theta\rightarrow\theta+\frac{2\pi}{n}$
in the expression of $c_{ml}$ we obtain $c_{ml}=e^{i(l-m)\frac{2\pi}{n}}c_{ml}$.
Therefore the coefficient must be zero for all $m$
except those that fulfill $l-m=kn$ with $k\in\mathbb{Z}$. On the other hand, $m$ must present a cut-off, in accordance with Th. \ref{thm:cut_off_en_m}.  This proves the following proposition:

\begin{prop}
\label{pro:regla_transf_vorticidad} Let us suppose that $V(\mathbf{x},t)$ is of the form given by Eq. (\ref{eq:cambio_V}) and that
$\phi(\mathbf{x},t\le t_{1})$ has well defined angular momentum $l$. Then $\phi(\mathbf{x},t\ge t_{1})$ has well defined angular pseudomomentum $m$ determined by the equation:
\begin{eqnarray}
l-m=kn,\nonumber
\end{eqnarray}
with  $k\in\mathbb{Z}$, $\left|m\right|\le\frac{n}{2}$ for even $n$  and $\left|m\right|\le\frac{n-1}{2}$ for odd $n$.
\end{prop}

\begin{figure}
\hspace{1.5cm}\includegraphics[scale=0.6] {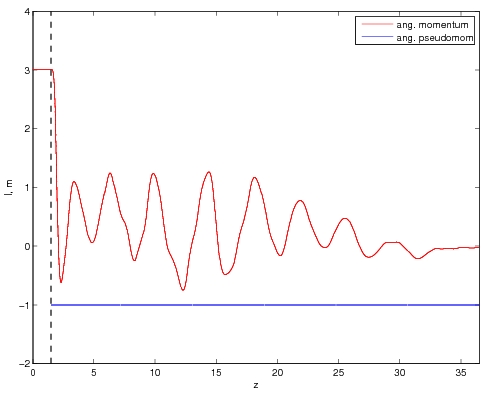}
\caption{\label{cap:l_and_m_vs_x} Evolution of the angular momentum and pseudomomentum when $l=3$ and $\mathcal{G}$ is $\mathcal{C}_{4v}$ from $t_1=1.5$.}
\end{figure}

\begin{figure} 
\begin{tabular}{cc}
 (a)&
\hspace{1.6cm}(b)\tabularnewline
\hspace{0.4cm}\includegraphics[%
  scale=0.45]{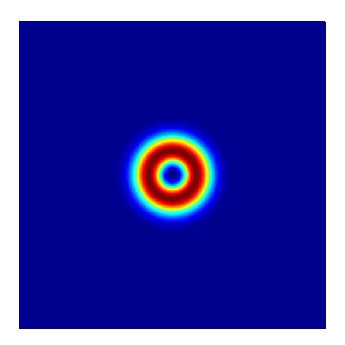}&
\includegraphics[%
  scale=0.45]{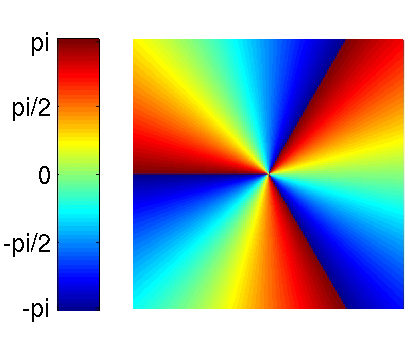} \tabularnewline
(c)&
\hspace{1.6cm}(d)\tabularnewline
\hspace{0.4cm}\includegraphics[%
  scale=0.45]{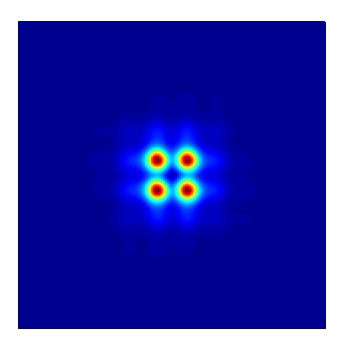}&
\includegraphics[%
  scale=0.45]{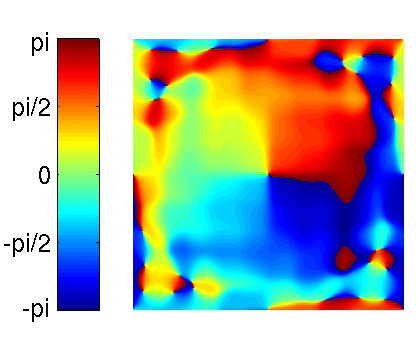} \tabularnewline
\end{tabular}
\caption{\label{cap:Amplitud-fase_O2-C4} Amplitude (a) and phase (b) of the solution at $t=t_1$ when $c=0.424$ and $r=2.1$, and amplitude (c) and phase (d) of the solution at $t=t_1+200$, when $l=3$ and $\mathcal{G}$ is $\mathcal{C}_{4v}$ from $t_1=1.5$.}
\end{figure}

Some examples of transformations satisfying this rule have been presented in Refs.~\cite{FerrandoPRL2005b,Perez_GarciaPRA07}. Let us analyze an example of a transformation from the point of view of angular momentum and pseudomomentum during the evolution. The function at $t=t_1$ is $\phi(\mathbf{x},t\le t_{1})=cr^{l}e^{-\frac{x^{2}+y^{2}}{r_{0}^{2}}}e^{il\theta}$ with $c=0.424$ and $r_{0}=2.1$. Therefore it  has angular momentum $l=3$. The function $V(\mathbf{x},t\ge t_1)$ is a squared lattice as in section \ref{c4v}. So, since  $m=l-kn$, $k\in\mathbb{Z}$, $n=4$, and $ \left|m\right|\le\frac{n}{2}$, necessarily $m=-1$. In Fig. \ref{cap:l_and_m_vs_x} we present the numerically calculated value of $l$ and $m$ during propagation, using  Eq. (\ref{eq:ang_momentum}) normalized by $\int dxdy\phi\phi^*$ to calculate $l$ and  discrete rotational transformations to calculate $m$.  As can be seen, $l$ for $t>t_1$ is no longer conserved but $m$ is.
If the definition of $j_{z}$ is applied to a
function with well defined angular pseudomomentum
$\phi(r,\theta,t)=e^{im\theta}u_{m}(r,\theta,t)$, with  $u_{m}(r,\theta+\epsilon,t)=u_{m}(r,\theta,t)$ the following result is obtained:
\begin{eqnarray}
j_{z}=m+\int_{\mathbb{R}^{2}}u_{m}^{*} \left(-i\frac{\partial}{\partial\theta}\right)u_{m}drd\theta \equiv m+j_{z}^{m}\,.\nonumber
\end{eqnarray}
So the conservation of  $m$ appears as a balance between
$j_{z}$ and  $j_{z}^{m}$. Note that $j_{z}^{m}$ is a function of time. 

In Fig. \ref{cap:Amplitud-fase_O2-C4} we present the amplitude and phase of $\phi(\mathbf{x},t= t_{1})$
and of the solution $\phi(\mathbf{x},t= t_{1}+200)$. As can be analyzed in Fig. \ref{cap:Amplitud-fase_O2-C4} (c) and (d) the solution belongs to the corresponding irreducible representation and has the proper angular pseudomomentum $m=-1$.

\section{Theory of Angular Pseudomomentum in the Quantum Limit}
\label{Theor}

Once the relevance of a proper treatment
of the angular pseudo-momentum in the description of discrete symmetry systems 
has been clearly established,
one question remains still open: are these properties still
valid in the limit where the mean-field approach is no longer
appropriate?  In other words, what would happen in the energy domain
where quantum effects start to be relevant?  Such a question, that at 
first glance could seem to be a purely academic one, has become extremely important 
since Bose-Einstein Condensates (BEC) became
experimentally feasible thirteen years ago~\cite{AndersonScience1995}. In order to provide
an answer to this question one should start by specifying a Hamiltonian
to describe the interactions at the bosonic level.  To do so we have
considered a standard interaction, the Bose-Hubbard Hamiltonian
\cite{JaksonPRL1998}. This model is a discretization of the full quantum many-body
Hamiltonian for weakly interacting ultracold bosons that in the
mean-field limit allow us to connect with the Gross-Pitaevskii
(nonlinear Schr\"odinger) equation~\cite{MishmasharXiv}. A complete analysis of the
 solution of this model within a periodic 1D bosonic chain has
been carried out~\cite{CarrInProgress}.  From these results it can be clearly observed
how the aforementioned angular pseudo-momentum relations also hold in
the case of few-boson systems once the angular pseudo-momentum is
properly identified with a quantum number by means of the generator of the symmetry group.

\section{Conclusions}
\label{Conc}

In conclusion, we have proved a number of theorems and propositions that show that the symmetric solitons of a generalized nonlinear Schr\"odinger equation can be classified according to the irreducible representations of the discrete rotational group of the modulus of their amplitude. We have define a new quantity, the \emph{angular pseudomomentum}, that can be used to classify nonlinear solutions in different representations. In this  way, the behaviour of any nonlinear symmetric solution under the discrete rotations of the group is completely determined by this number. Moreover, we have demonstrated several theorems about the allowed values of the angular pseudomomentum. We have illustrated these mathematical results with numerical examples in $\mathcal{C}_{4v}$ and $\mathcal{C}_{6v}$. We have also extended these results to non-stationary solutions and we have proved a theorem that states  that the angular pseudomomentum is  dynamically conserved.  We have  demonstrated how angular momentum and pseudomomentum can be related in a $\mathcal{O}(2)-\mathcal{C}_{n}$ interface.  We have explained that the non-conservation of angular momentum in $\mathcal{C}_{nv}$  groups  appears as a balance between the angular pseudomomentum and another time-dependent quantity $j_z^m$ related to the angularly periodic part ($u(r,\theta,z )$) of the solution.  We have presented a numerical example to illustrate these dynamical results. Finally, we have outlined the generalization of these results to the quantum limit which is relevant to trapped atoms in an optical lattice when a very low number of atoms is considered. It is expected that the most important results of this paper survive in this limit since they are preserved by symmetry arguments that should be impervious to dynamical details, but this remains an open question at present.

LDC acknowledges support by the National Science
Foundation under Grant PHY-0547845 as part of the NSF CAREER program.


\end{document}